\documentclass{article}

\usepackage{microtype}
\usepackage{graphicx}
\usepackage{subfigure}
\usepackage{booktabs} 
\usepackage[inline]{enumitem}
\usepackage{multirow}
\usepackage{xcolor}
\usepackage{xspace}

\usepackage{hyperref}

\usepackage[accepted]{mlsys2024}

\newcommand{\sys}{LLM microserving\xspace}
\newcommand{\papertitle}{A System for Microserving of LLMs}

\usepackage[normalem]{ulem}
\definecolor{applegreen}{rgb}{0.55, 0.71, 0.0}
\mlsystitlerunning{\papertitle}

\begin{document}

\newcommand{\keypointcomment}[1]{}

\newcommand{\vspacebeforecap}{\vspace{-2em}}
\newcommand{\vspaceaftercap}{\vspace{-1em}}


\newcommand{\MyPara}[1]{\vspace{.0em}\noindent\textbf{#1}}

\newcommand{\squishlist}{
   \begin{list}{$\bullet$}
    { \setlength{\itemsep}{1pt}      \setlength{\parsep}{3pt}
      \setlength{\topsep}{3pt}       \setlength{\partopsep}{0pt}
      \setlength{\leftmargin}{1em} \setlength{\labelwidth}{1em}
      \setlength{\labelsep}{0.5em} } }

\newcommand{\squishlisttwo}{
   \begin{list}{$\bullet$}
    { \setlength{\itemsep}{0pt}    \setlength{\parsep}{0pt}
      \setlength{\topsep}{0pt}     \setlength{\partopsep}{0pt}
      \setlength{\leftmargin}{2em} \setlength{\labelwidth}{1.5em}
      \setlength{\labelsep}{0.5em} } }

\newcommand{\squishend}{
    \end{list}  }

\twocolumn[
\mlsystitle{\papertitle}


\mlsyssetsymbol{equal}{*}

\begin{mlsysauthorlist}
\mlsysauthor{Hongyi Jin}{equal,cmu}
\mlsysauthor{Ruihang Lai}{equal,cmu}
\mlsysauthor{Charlie F. Ruan}{equal,cmu}
\mlsysauthor{Yingcheng Wang}{equal,tsinghua}
\mlsysauthor{Todd C. Mowry}{cmu}
\mlsysauthor{Xupeng Miao}{purdue}
\mlsysauthor{Zhihao Jia}{cmu,amazon}
\mlsysauthor{Tianqi Chen}{cmu,nvidia}
\end{mlsysauthorlist}

\mlsysaffiliation{cmu}{Carnegie Mellon University}
\mlsysaffiliation{nvidia}{NVIDIA}
\mlsysaffiliation{amazon}{Amazon}
\mlsysaffiliation{tsinghua}{Tsinghua University}
\mlsysaffiliation{purdue}{Purdue University}

\mlsyscorrespondingauthor{Hongyi Jin}{hongyij@cs.cmu.edu}

\mlsyskeywords{Machine Learning, MLSys}

\vskip 0.3in

\begin{abstract}

The recent advances in LLMs bring a strong demand for efficient system support to improve overall serving efficiency. As LLM inference scales towards multiple GPUs and even multiple compute nodes, various coordination patterns, such as prefill-decode disaggregation and context migration, arise in serving systems. 
Most inference services today expose a coarse-grained request-level API with a pre-configured coordination strategy, limiting the ability to customize and dynamically reconfigure the coordination. In this paper, we propose LLM microserving, a multi-level architecture for structuring and programming LLM inference services. We introduces simple yet effective microserving APIs to support  fine-grained sub-request level actions. A programmable router transforms user requests into sub-request calls, enabling the dynamic reconfiguration of serving patterns. To support diverse execution patterns, we develop a unified KV cache interface that handles various KV compute, transfer, and reuse scenarios. Our evaluation shows that LLM microserving can be reconfigured to support multiple disaggregation orchestration strategies in a few lines of Python code while maintaining state-of-the-art performance for LLM inference tasks. Additionally, it allows us to explore new strategy variants that reduce up to 47\% of job completion time compared to the existing strategies.

\end{abstract}
]

\printAffiliationsAndNotice{}  

\section{Introduction}
\label{sec:introduction}

Large language models (LLMs) have achieved remarkable capabilities, handling diverse tasks like text generation~\cite{touvron2023llama2openfoundation}, question answering, and code synthesis~\cite{rozière2024codellamaopenfoundation}.
The recent advances in LLMs bring a strong demand for efficient system 
support for improving the overall serving efficiency. 
As LLM serving scales towards multiple GPUs and even multiple compute instances, many coordination and optimization patterns  arise. 
For example, prefill-decode disaggregation~\cite{distserve, patel2024splitwise} separates the prefilling and decoding stages of LLM serving and assigns them to dedicated GPU instances.
As we start to bring prefix caching and radix attention~\cite{sglang} across serving instances, there is also a strong demand to enable effective migration of the cached prefix to reduce overall redundant computation across multiple requests in the system. 
There are also multiple ways to dispatch an input request across different LLM workers based on the ongoing
traffic~\cite{DBLP:conf/osdi/SunHZXZL024}. The complex combination of the coordination opportunities creates a rich space for orchestrating LLM deployment on multiple devices.

Most of the LLM inference services today build on top of systems that implement a fixed set of the strategies
and expose a request-level REST API for the end users. The configuration of the coordination strategies
is hidden behind the request-level endpoint and managed by the underlying system. Changing the strategy usually involves reconfiguring the underlying system and restarting the service, causing disruptions to the production environment. The coarse-grained LLM serving architecture also limits our ability to customize and explore different 
coordination strategies and dynamically reconfigure them based on the incoming traffic.

\begin{figure*}[!t]
    \centering
    \includegraphics[width=1\textwidth]{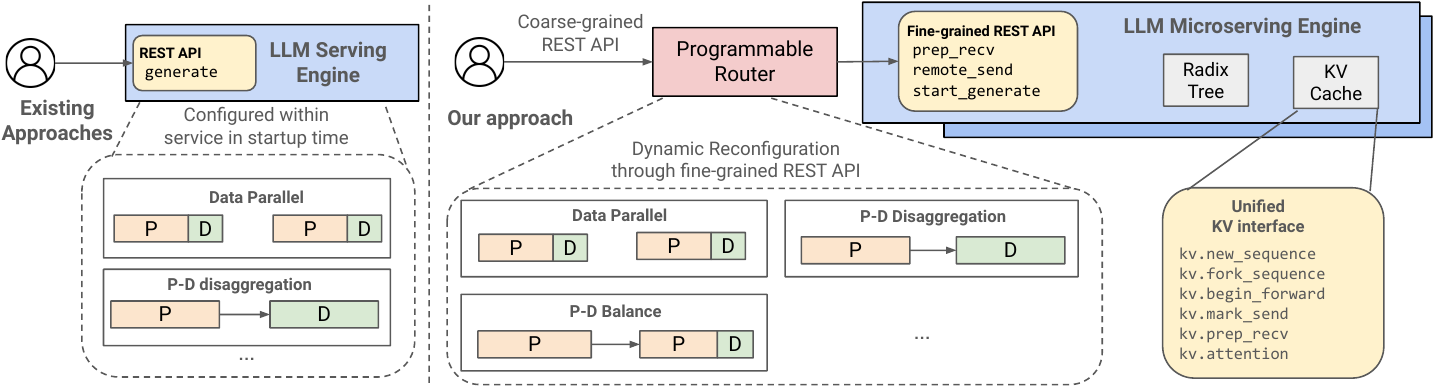}
    \caption{\sys{} System Overview. Our architecture enables dynamic reconfiguration of different orchestration strategies with a programmable router through three fine-grained REST APIs. The LLM microserving engines implement the APIs with a unified KV cache interface.}
    \vspaceaftercap
    \label{fig:overview}
\end{figure*}

This paper aims to address the above gap by introducing a multi-level architecture 
for structuring and programming LLM inference services.
We propose \sys{}, which exposes simple yet effective APIs for fine-grained sub-request 
level actions.
At the global level, we introduce a programmable router that transforms an end-user-provided API request into corresponding calls to microserving endpoints using python async functions.
Such transformation enables flexible program schedules that orchestrate the LLM engines on the fly, enabling dynamic reconfiguration according to the dynamic workloads.
Finally, we build a unified KV cache abstraction to support the common attention transfer,
reuse, and compute patterns, and use that to build an efficient microserving engine for 
LLM inference.

The proposed microserving architecture enables developers to easily specify multiple
disaggregation and coordination patterns dynamically in the router through a simple yet effective
asynchronous Python programming model. 
For example, \sys{} can easily reproduce existing static scheduling patterns, such as data parallel and prefill-decode disaggregation.
Dynamic scheduling patterns such as prefill-decode work balancing and distributed context cache management can also be supported by programming with the REST APIs.

We implement the proposed abstractions in an end-to-end LLM inference engine. Our evaluation
demonstrates that \sys{} enables a flexible programming model to support multiple disaggregation and orchestration 
strategies while maintaining state-of-the-art performance for LLM inference tasks.
The flexible programming model also enables us to quickly explore a new disaggregation strategy
that pushes part of prefill workloads to decode for load balancing. Our evaluation shows
that the balanced prefill-decode disaggregation strategy can reduce up to 47\% job completion time compared to existing strategies.

This paper makes the following contributions:

\begin{itemize}
    \item We propose a microserving architecture for LLM inference service with simple yet effective composable fine-grained APIs.
    \item We introduce a programmable router to enable dynamic reconfiguration of multiple LLM inference orchestration patterns.
    \item We build a unified KV interface that handles model computation under different KV transfer,
    reuse, and compute patterns for microserving.
\end{itemize}

\section{Overview}
\label{sec:overview}

\sys{} is an architecture that enables dynamic reconfiguration of different disaggregation and coordination patterns. It allows systems to smoothly adapt to the traffic during production, and offers framework users fine-grained control over scheduling their systems.

As depicted in Figure \ref{fig:overview}, existing frameworks implement a fixed set of strategies encapsulated in the underlying LLM serving engine and expose a coarse-grained request-level API to developers. As a result, reconfiguration may require system restarts and disrupt the production environment, and framework users have limited abilities to schedule their system dynamically according to the traffic. \sys{} architecture addresses this gap with three key components.

First of all, \sys{} architecture defines three simple fine-grained APIs that allow framework users to express various orchestration patterns when composed together (\S\ref{sec:rest_api}).

In the control layer, the programmable router transforms the request-level API into the fine-grained APIs dispatched to the engines. The router's simple asynchronous Python programming model allows framework users to program their own router to dynamically specify different disaggregation and coordination patterns, including data parallel, prefill-decode disaggregation, context cache migration, and more (\S\ref{sec:router}). It also enables exploration of new strategy variants that may reduce up to 47\% of job completion time compared to existing strategies (\S\ref{sec:balanced_pd_disagg}).

In the execution layer, to support the diverse semantics of the fine-grained APIs, the same LLM microserving engines need to organically execute a combination of different KV transfer, reuse, and compute patterns. To achieve this, we propose a unified KV interface that abstracts out such execution-level patterns (\S\ref{sec:engine_kv}). To minimize KV transfer overhead, the engines implement KV transfer with async and one-sided GPU communication (\S\ref{sec:e2e_impl}).

With these components, \sys{} enables a flexible programming model to orchestrate the underlying system dynamically, while maintaining state-of-the-art performance for LLM inference tasks (\S\ref{sec:evaluation}).

\section{Our Approach}
\label{sec:approach}
\subsection{Microserving REST API} \label{sec:rest_api}
Existing LLM serving systems typically treat token generation for a request as an atomic operation, requiring developers to implement different code inside the underlying system for each scheduling pattern. This approach limits framework users' abilities to schedule their system dynamically, and it often necessitates system restarts when reconfiguring the scheduling pattern, causing disruptions in production environments. Our insight is that common scheduling patterns can be expressed using two fundamental actions:
\begin{itemize}
    \item Transferring some KV cache from one LLM engine to another. The transferred KV may originate from either the existing context cache or the new prefill computation.
    \item Initiating token generation with full or partial KV cache of the prompt.
\end{itemize}

Since transferring KV requires coordination between sender and receiver, we introduce three fine-grained REST APIs (Figure \ref{table:rest_api}) microserving the actions. By invoking these APIs from the router, developers can compose various scheduling patterns and easily reconfigure the system from one pattern to another.
\begin{table}[t]
    \centering
    \caption{Three fine-grained REST APIs}
    \includegraphics[width=0.5\textwidth]{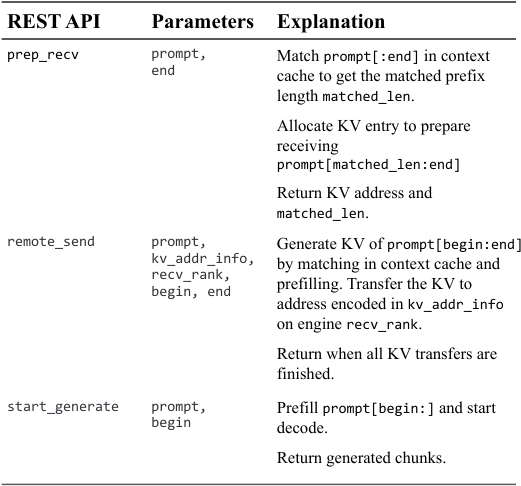}
    \vspaceaftercap
    \label{table:rest_api}
\end{table}

\texttt{prep\_recv} checks for the existence of \texttt{prompt[:end]} in the KV cache. It allocates KV cache entries for the missing subsequence and returns their address in a compressed form.

\texttt{remote\_send} sends KV of \texttt{prompt[begin:end]} to engine \texttt{recv\_rank}. If part of the subsequence already exists in the KV cache, it is directly transferred to the receiver. For other parts, new KVs are first materialized through prefill computation and then sent to the receiver.

\texttt{start\_generate} informs the engine that all KVs of \texttt{prompt[:begin]} already exist in its KV cache and it's ready to prefill \texttt{prompt[begin:]} and start decoding.

By calling the APIs sequentially, we can create a workflow like the following. The router wants to perform partial prefill and decode on engine A. It first determines which part of KV is missing on engine A (with \texttt{prep\_recv}), and then asks engine B to transfer those missing parts to engine A (with \texttt{remote\_send}). After engine A receives all necessary KV, it proceeds with the remaining prefill and begins decoding (with \texttt{start\_generate}). Different compositions of the APIs can support more patterns as discussed below.

\subsection{Programmable Router} \label{sec:router}

With fine-grained REST APIs enabling granular control over engine actions, the router can transform a coarse-grained request-level API into a set of fine-grained REST APIs based on the current scheduling pattern.

Reconfiguration between scheduling patterns only occurs on the router side and does not require engine reconfiguration.

We showcase several examples of how the router can be programmed to support various scheduling patterns using concise asyncio Python code. These examples can be generalized to a router that dynamically reconfigures its strategy by inspecting an internal policy updated by the traffic.

\begin{figure}[t]
    \centering
    \includegraphics[width=0.5\textwidth]{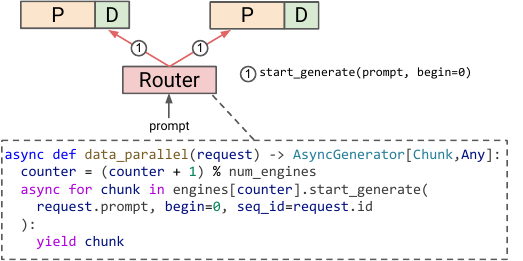}
    \caption{Data parallel via microserving}
    \vspaceaftercap
    \label{fig:visualize-dp}
\end{figure}

\paragraph{Example 1: Data parallel} Data parallel (Figure~\ref{fig:visualize-dp}) is the strategy that dispatches incoming requests to engines in a round-robin fashion. It is straightforward to implement because engines do not communicate with each other, making it the default strategy in many LLM serving systems. While data parallel helps achieve high throughput, it does not reduce latency except for the reduced decode batch size.

To implement data parallel in the router, we only need to maintain a counter indicating the next engine to dispatch the request to, and then send a \texttt{start\_generate(prompt,begin=0)} API call to that engine. Figure~\ref{fig:visualize-dp} contains a reference router implementation of data parallel in just 5 lines of code.

\paragraph{Example 2: Prefill-decode disaggregation} Prefill-decode disaggregation separates prefill and decode operations across LLM engines. The key idea is that colocating prefill and decode may result in strong interference and unnecessary coupling of resource allocation and parallelism between the two phases.  Disaggregating prefill and decode leads to lower latency, better resource management, and more scalability. This scheduling strategy is challenging to implement in two aspects: efficiently transferring KV cache from the prefill engine to the decode engine, and coordinating between the two engines to minimize CPU runtime overhead. Our fine-grained REST API provides a clean and easy-to-use solution to the coordination issue, while our unified KV cache interface enables asynchronous and fast KV transfer (\S\ref{sec:e2e_impl}).

\begin{figure}[t]
    \centering
    \includegraphics[width=0.48\textwidth]{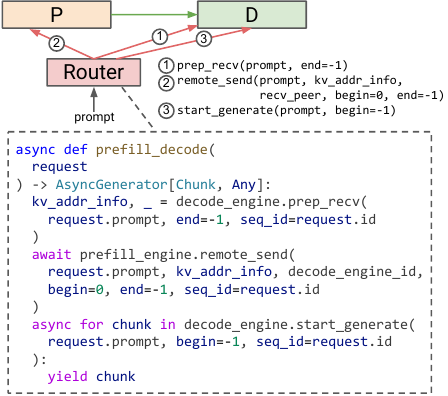}
    \caption{Prefill-decode disaggregation via microserving}
    \vspaceaftercap
    \label{fig:visualize-disagg}
\end{figure}

To illustrate the router-side implementation, let's consider a simple case where we do not have a context cache. As shown in Figure \ref{fig:visualize-disagg}, the router first calls \texttt{prep\_recv} on the decode engine. The argument \texttt{end=-1} instructs the decode engine to prepare space for the KV cache entries of \texttt{request.prompt[:-1]} (all but the last prompt token), and the returned \texttt{kv\_addr\_info} encodes the page index and slot index of the KV cache entries.

Next, the router calls \texttt{remote\_send} on the prefill engine, with \texttt{begin=0} and \texttt{end=-1} to ask the prefill engine to run prefill computation to generate KV cache for \texttt{request.prompt[:-1]} and send KV cache to the corresponding entries on the decode engine side. The KV transfer here uses a remote memory access semantic, allowing the sender to directly write KV into the remote address.  When the remote write is completed \texttt{remote\_send} returns.

After \texttt{remote\_send} returns, the KV transfer is complete. The router calls \texttt{start\_generate} on the decode engine to prefill the last token, sample, and start decoding.

Figure \ref{fig:visualize-disagg} shows an API call graph illustrating these three steps and their implementation. The scheduling patterns discussed later have similar router-side implementations to prefill-decode disaggregation, differing only in the arguments of the APIs.

\paragraph{Example 3: Context cache-aware prefill-decode disaggregation}
This example extends the prefill-decode disaggregation to consider the context cache on each LLM engine. The implementation is similar to the no-context-cache case, with the main difference being that \texttt{match\_len} is not necessarily 0 (Figure \ref{fig:visualize-context-pd-disagg}).

\begin{figure}[t]
    \centering
    \includegraphics[width=0.48\textwidth]{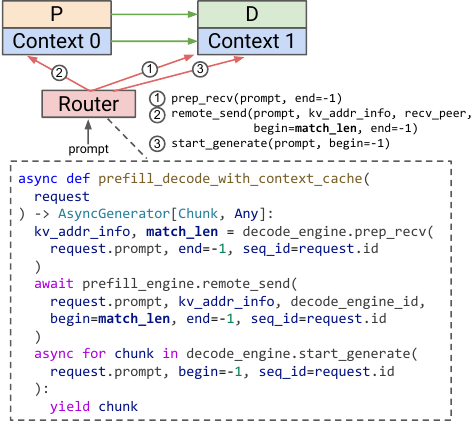}
    \caption{Context cache-aware prefill-decode disaggregation via microserving}
    \vspaceaftercap
    \label{fig:visualize-context-pd-disagg}
\end{figure}
The key API call differences are:
\squishlist
    \item \texttt{prep\_recv}: The decode engine matches the prompt with its local context cache, returning the matched prefix length \texttt{match\_len} and allocating space for the unmatched portion.
    \item \texttt{remote\_send}: The prefill engine matches the prompt with its local context cache and transfers the necessary KV data to the decode engine. This may involve direct transfer of cached data and/or prefill computation for unmatched portions.
    \item \texttt{start\_generate}: Remains the same as in the no-context-cache case.
\squishend

Therefore, the REST APIs allow efficient handling of different prefix-matching scenarios between the prefill and decode engines. For more details on the internal workings of KV transfer in various cases, refer to \S\ref{sec:kv_context}.

\paragraph{Example 4: Context Cache Migration}

When serving QA workloads, developers tend to place context cache of different categories into different engines and dispatch incoming traffic based on which context category it matches. Consider there are several engines, with some specialized for history context and others for science context. If there are more science requests than history requests, we may want to switch some history engines to science engines through context migration, and vice versa.

\begin{figure}[t]
    \centering
    \includegraphics[width=0.47\textwidth]{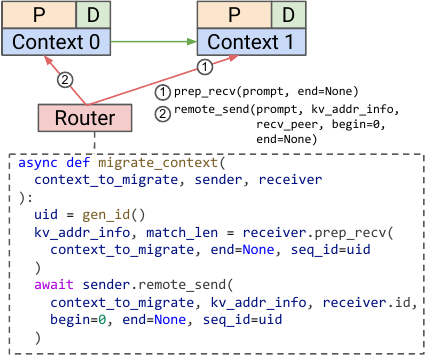}
    \caption{Context cache migration via microserving}
    \vspaceaftercap
    \label{fig:visualize-context-migration}
\end{figure}

Context cache migration can be composed with our fine-grained REST APIs as well (Figure \ref{fig:visualize-context-migration}). By calling \texttt{prep\_recv} on receiver and \texttt{remote\_send} on sender, a context transfer is easily accomplished. Note that the router also needs to maintain a radix tree in addition to those on LLM engines, so that it can decide which engine to dispatch requests to and when to trigger a context switch.

\subsection{Exploring Balanced Prefill-Decode Disaggregation} \label{sec:balanced_pd_disagg}
\begin{figure}[t]
    \centering
    \includegraphics[width=0.48\textwidth]{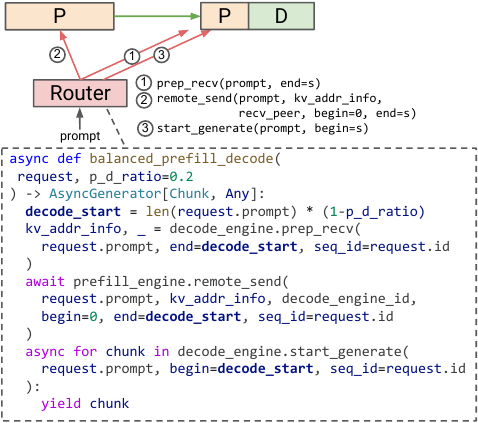}
    \caption{Balanced prefill-decode disaggregation via microserving}
    \vspaceaftercap
    \label{fig:visualize-pd-balance}
\end{figure}

\S\ref{sec:router} shows \sys's functionality to leverage the router programming model for supporting existing strategies for disaggregating inference and context migration. 
This section introduces how this programming flexibility enables \sys to support new prefill-decode disaggregation strategies that haven't been considered by prior work. One issue of prefill-decode disaggregation is that the prefill and decode workloads can get unbalanced depending on the workload~\cite{qin2024mooncake}. 
When processing long prompts, the prefill engine can get over-utilized while the decode engine stays idle. Most prior approaches~\cite{hu2024inferenceinterferencedisaggregatellm,jin2024pdserveservingdisaggregatedlarge,wu2024loongserve} solve this problem by dynamically adapting the number of prefill and decode instances, which introduces significant migration overheads. 

We explore a different strategy that {\em dynamically} moves a part of prefill computation into the decode engine, where the migrated prefill computation can be fused with the decode computation to improve overall throughput. We refer to this strategy as {\em balanced prefill-decode disaggregation}, and implement it in a few lines of code using the router's programming interface (see \autoref{fig:visualize-pd-balance}). 
The router decides the subsequence (\texttt{prompt[:s]}) to compute on the prefill instance based on the monitored P:D distribution. When the prefill engine returns from \texttt{remote\_send}, the decode engine has KV of \texttt{prompt[:s]}, so it continues to prefill the remaining part of \texttt{prompt[s:]}, does sampling, and starts decoding. This strategy enables more fine-grained load balancing between prefill and decode, making disaggregation applicable to a broader range of workloads.

\subsection{Microserving with a Unified KV Cache Interface} \label{sec:engine_kv}
\begin{figure*}[t]
    \centering
    \includegraphics[width=1\textwidth]{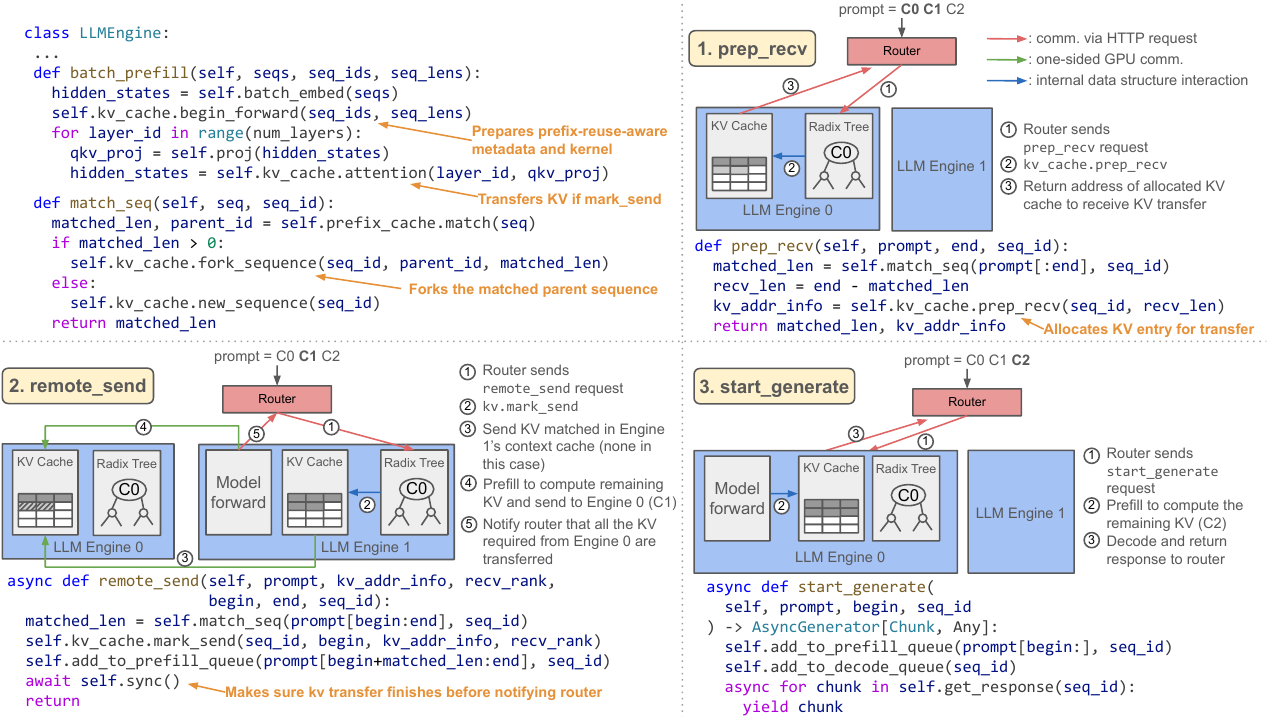}
    \vspacebeforecap
    \caption{The implementation of each REST API in an LLM engine with a unified KV Cache interface. Here we depict prefilling a prompt with contexts C0 C1 C2 in a balanced P-D disaggregated pattern discussed in \S\ref{sec:balanced_pd_disagg}. Engine 0 is the decode engine, and Engine 1 is the prefill engine. The router instructs Engine 0 to prefill C2 to transfer part of prefill pressure. Both engines have C0 in their context cache. (1) The router sends \texttt{prep\_recv} to Engine 0, which matches C0 and tells the KV cache to prepare for receiving C1. (2) The router sends \texttt{remote\_send} to Engine 1. Engine 1 also matches C0, prefills C1, and sends the KV of C1 to Engine 0 with remote GPU memory write. (3) After the KV transfer of C1 finishes, the router sends \texttt{start\_generate} to Engine 0, which prefills C2 first and then starts decoding.}
    \vspaceaftercap
    \label{fig:engine_impl}
\end{figure*}

\begin{table}[t]
    \centering
    \caption{The KV Cache interface.}
    \includegraphics[width=0.5\textwidth]{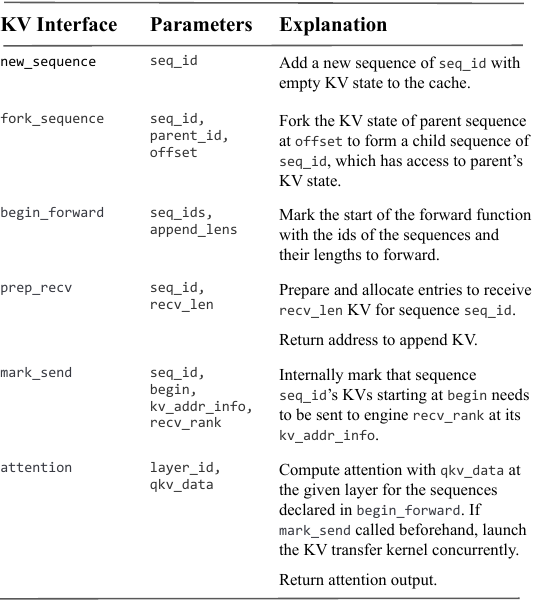}
    \vspaceaftercap
    \vspace{-1.5em}
    \label{table:kv_api}
\end{table}

The previous sections discuss the control plane by defining the semantics of the three fine-grained APIs and showing how they can schedule the system dynamically. This section goes over how an LLM microserving engine can implement and execute each of the three fine-grained APIs with a unified KV cache interface, as shown in Figure \ref{fig:engine_impl}. The KV cache APIs are listed in Table \ref{table:kv_api}.

To implement the diverse semantics of the REST APIs, the same engine needs to support various attention compute, KV transfer, and KV reuse patterns. Under prefill-decode disaggregation, the engine needs to send part of the KV cache to the remote; while prefix matching can result in forking operations where multiple sequences share the same prefix, allowing the attention operation to load the KV of the shared prefix once. These may also result in different combinations (\S\ref{sec:kv_context}). Therefore, it is challenging to support the numerous patterns in the same model definition.

Our unified KV cache interface is yet another simple and effective solution for abstracting out such patterns. It provides a two-stage interface:

\squishlist
    \item In the \emph{declaration} stage, the KV cache provides two main runtime APIs to declare the necessary information about the pending actions. \texttt{begin\_forward} declares pending attention operations and asks the KV cache to \emph{plan} the set of metadata and kerne kernel operations once for all attention layers. \texttt{mark\_send} declares that the pending attention would invoke a KV transfer to remote.
    \item In the \emph{computation} stage during model forward, the model definition calls \texttt{attention} with \texttt{qkv\_data}. The KV cache performs attention with the already-prepared metadata, which informs the KV cache of outstanding KV transfers and opportunities to leverage common prefixes.
\squishend

With such a two-stage KV interface, the engine can implement each of the three fine-grained REST APIs. We go over each implementation while referring to Figure \ref{fig:engine_impl}.

To implement \texttt{prep\_recv}, the engine calls the KV cache API \texttt{prep\_recv} that allocates KV entries to prepare for receiving the unmatched parts of the prompt and returns the address that the peer should send to. There is no computation for this REST API.

For \texttt{remote\_send}, the engine calls the KV cache API \texttt{mark\_send} to declare that KV transfer is needed for this sequence. The engine then uses \texttt{begin\_forward} to inform the KV cache of any prefix reuse opportunities for the upcoming compute. Upon \texttt{attention} in model forward, the KV cache sends this sequence's KVs to the peer via GPU communication, overlapped with attention computation.

Upon the execution of \texttt{start\_generate}, the engine already possesses the KVs both from its local KV cache and from the remote peer (if any). It uses the same \texttt{begin\_forward} and \texttt{attention} steps to prefill any remaining part of the prompt and starts decoding.

\subsection{KV Cache and Distributed Context Cache} \label{sec:kv_context}

To support prefix matching as promised by the REST APIs, the LLM engines need to maintain a context cache. While some existing frameworks implement this inside the KV cache, our KV cache API \texttt{fork\_sequence} decouples context cache management from the KV cache. This allows for the coexistence of both local eviction policy from the engine and global eviction policy from the router, which may instruct the engine to ``pin'' certain important prefixes based on its global knowledge.

When \texttt{mark\_send} is called, \texttt{attention} is responsible for sending the KVs to the receiver. However, this process is complicated by the possibility of prefix reuse. For instance, a part of the KVs to be sent may already exist in the sender's context cache, allowing direct transfer to the receiver. Here, we discuss KV cache transfer in different prefix-matching scenarios by considering two cases (Figure \ref{fig:kv_prefix_match}).

\begin{figure}[t]
    \centering
    \includegraphics[width=0.48\textwidth]{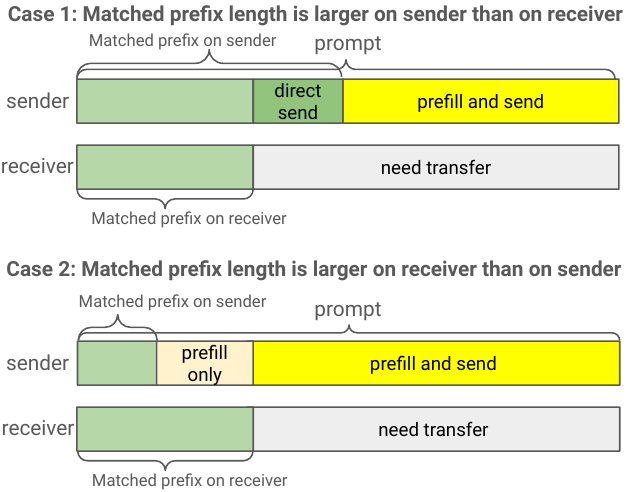}
    \vspacebeforecap
    \vspace{1em}
    \caption{\texttt{attention} handles KV transfer in different prefix-matching scenarios.}
    \vspaceaftercap
    \label{fig:kv_prefix_match}
\end{figure}

\textbf{Case 1:} Matched prefix length is larger on sender than on receiver. In this scenario, the segment matched only by the sender (in dark green) can be directly sent to the receiver, while the segment matched by neither engine needs to be prefilled and then transferred (in yellow).

\textbf{Case 2:} Matched prefix length is larger on receiver than on sender. In this scenario, not all the KV segments needed for transfer exist in the sender's context cache. To compute the segment, the sender should prefill all the non-cached segments and transfer only the part required by the receiver.

Given the numerous possible combinations, it is crucial to have a unified interface that supports different KV transfer, reuse, and compute patterns. 

\subsection{End-to-end Implementation} \label{sec:e2e_impl}

We implement the LLM microserving engine and KV cache on top of the existing project MLC-LLM~\cite{mlc-llm}, which has a total of 13k lines of C++ and 6k lines of Python. Thanks to the flexibility of \sys{}, we implement all the strategies in the router with \emph{350 lines} of Python code.

For GPU communication, we use the NVSHMEM library~\cite{nvshmem} which, unlike NCCL, supports one-sided \texttt{put} and \texttt{get} API. That is, the communication only requires SM on one side to participate. We overlap the KV transfer with attention computation by having a computation stream and a communication stream and by eagerly sending a layer's KV right after its computation finishes, as shown in Figure~\ref{fig:overlap}. The KV transfer is asynchronous and does not affect the ongoing requests of the receiver engine.

\begin{figure}[t]
    \centering
    \includegraphics[width=0.5\textwidth]{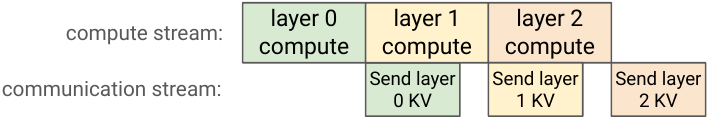}
    \vspace{-1em}
    \caption{Overlapping of KV transfer and attention computation.}
    \vspaceaftercap
    \label{fig:overlap}
\end{figure}

\section{Evaluation}
\label{sec:evaluation}
This section provides an evaluation to answer the following questions:
\squishlist
    \vspace{-0.5em}
    \item Can we easily program customized LLM disaggregated serving patterns with \sys{}~(\S\ref{sec:eval:disagg-patterns})?
    \vspace{-0.2em}
    \item Can \sys{} support global context reuse and efficient KV migration between engines~(\S\ref{sec:eval:migration-efficiency})?
    \vspace{-0.2em}
    \item What are the performance impacts of PD balance ratio~(\S\ref{sec:eval:balance-ratio})?
\squishend

\subsection{Disaggregated LLM Inference Pattern Evaluation}
\label{sec:eval:disagg-patterns}

This section evaluates \sys{}'s programmability for various LLM disaggregated
serving patterns and the performance of each pattern.
We implement three disaggregation patterns, including
\squishlist
    \vspace{-0.5em}
    \item 1P1D: prefill-decode disaggregation with 1 prefill engine and 1 decode engine ~(Figure~\ref{fig:visualize-disagg}).
    \vspace{-0.2em}
    \item 1P1D-balance: balanced prefill-decode disaggregation with 1 prefill engine and 1 decode engine~(Figure~\ref{fig:visualize-pd-balance}).
    \vspace{-0.2em}
    \item 1P2D: prefill-decode disaggregation with 1 prefill engine and 2 decode engines.
\squishend
Notably, the programmability of \sys{} allows for sharing the same engine implementation across these patterns,
and the patterns only differ in how the router is implemented.
We also include the data parallelism (DP) implementation in \sys{} (Figure~\ref{fig:visualize-dp}) and vLLM~(v0.6.3.post1)~\cite{vllm} as baselines,
where vLLM runs with optimized latency configuration of 10-step scheduler.
Each engine runs the Llama3.1 8B model on a single NVIDIA A100 SXM 40GB GPU.
We use the balance ratio 0.2 for 1P1D-balance,
which means the prefill of the last 20\% tokens is assigned to the decode engine,
and the prefill engine only computes and transfers KV for the first 80\% of tokens.
The request arrival follows the arrival time sampled from the Poisson distribution of different per-GPU request rates,
in order to fairly compare disaggregation patterns that use different numbers of GPUs.
For each request, we measure the metrics of time to first token (TTFT), time per output token (TPOT), and job completion time (JCT).

We study the system performance under the ShareGPT dataset and synthetic datasets.
ShareGPT is a collection of user-shared conversations with ChatGPT and provides diverse examples of real interactions.
It features short input and output lengths, with means being around 200 and 260 respectively.
To study the performance with longer input lengths,
we generate the synthetic dataset with the input and output lengths of each entry sampled from a normal distribution.
The mean input length is 3000, and the mean output length is 100, with a standard deviation of 5 for both.
This configuration ensures that requests will not finish shortly after prefill, as typical QA datasets have few output tokens.

\begin{figure}[t]
    \centering
    \includegraphics[width=0.48\textwidth]{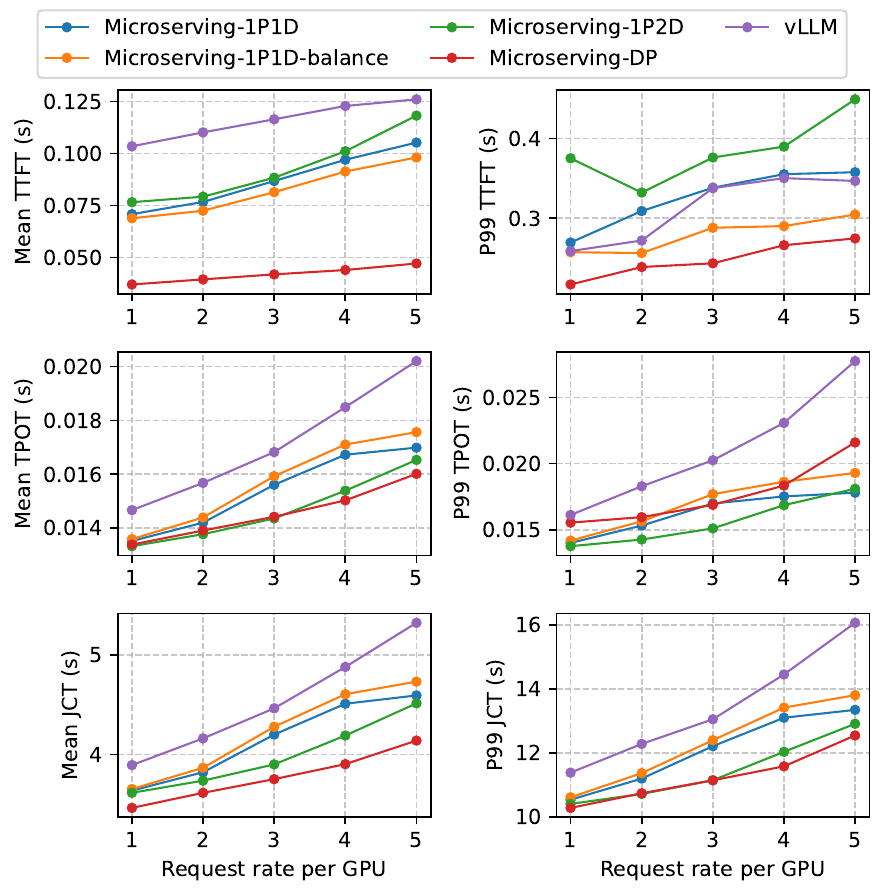}
    \caption{LLM inference evaluation on ShareGPT. Prefill-decode disaggregation has no observed benefit over data parallelism because the prefill engine is idle with the short input in ShareGPT.}
    \vspaceaftercap
    \vspace{1.2em}
    \label{fig:eval:sharegpt}
\end{figure}

\begin{figure}[t]
    \centering
    \includegraphics[width=0.48\textwidth]{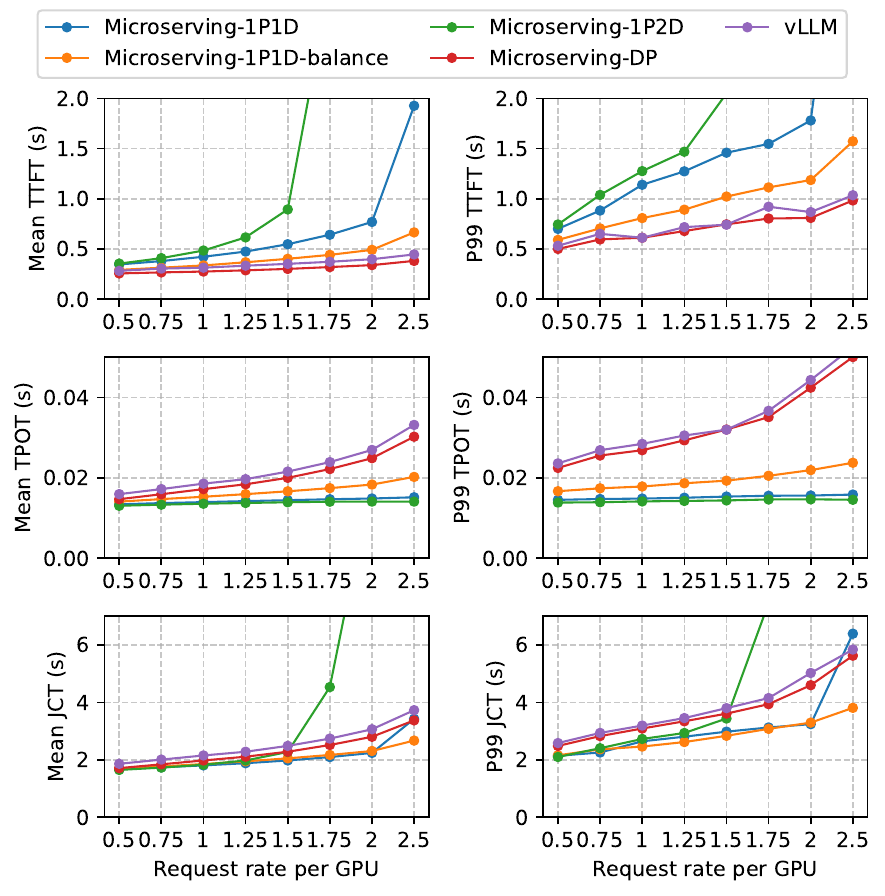}
    \caption{LLM inference evaluation on synthetic data with average input length 3000 and average output length 100. The new disaggregation pattern 1P1D-balance reduces up to 47\% of job completion time, which benefits from transferring part of prefill engine pressure to the decode engine.}
    \vspaceaftercap
    \label{fig:eval:synthetic}
\end{figure}

Figure~\ref{fig:eval:sharegpt} shows the evaluation results with the ShareGPT dataset.
When the requests follow ShareGPT input and output length distribution, data parallel remains a strong baseline overall, and we do not observe benefits of prefill-decode disaggregation.
This is mainly because the ShareGPT dataset features the same magnitude of input and output
lengths, so the prefill load is relatively low,
causing the prefill engine not to be fully utilized, while the data parallel engine is always kept busy.
Among the disaggregation patterns, 1P1D-balance underperforms the vanilla 1P1D due to the same reason of the input-output length ratio,
as 1P1D-balance further reduces the prefill engine workload and increases the decode engine workload.
1P2D brings lower TPOT and JCT than 1P1D by doubling the decode engine and amortizing the decode pressure onto two engines.

Figure~\ref{fig:eval:synthetic} shows the evaluation results of the synthetic dataset, which features longer input lengths that increase the prefill engine workload.
Prefill-decode disaggregation demonstrates the capability of reducing the JCT compared to data parallelism by up to 21\% for mean JCT and up to 47\% for P99 JCT.
This significant speedup attributes to the substantial reduction of TPOT in disaggregation, achieved by eliminating the decode interference caused by long prefill in data parallelism.
As the per-GPU request rate increases to 2.5 req/s, the heavier traffic puts more pressure on the prefill engine and causes performance degradation in 1P1D due to the increase of TTFT.
In this case, the 1P1D-balance pattern helps by transferring some of the prefill engine pressure to the decode engine, thus maintaining a low JCT.

The disaggregated LLM inference pattern study shows that different patterns have different preferences on request input-output length ratios and traffic,
and striking a workload balance between the prefill and decode engines is a key to better performance.
\sys{} provides microserving APIs and programmability, supporting the representation of various disaggregation patterns and the dynamic reconfiguration among these patterns.

\subsection{KV Migration Efficiency Evaluation}
\label{sec:eval:migration-efficiency}

In this section, we evaluate the efficiency of KV migration from one LLM engine \texttt{E1} to another engine \texttt{E2} when context cache is available~(Figure~\ref{fig:visualize-context-pd-disagg}).
The evaluation method is: First prefill a context on \texttt{E1}. Then create a new prompt by concatenating the context with unique text of 500 tokens, and run its prefill on \texttt{E1} and decode on \texttt{E2}.
 When processing the prefill, \texttt{E1} recognizes the context reuse by context cache match, and then automatically migrates the context KV data to \texttt{E2}. \texttt{E1} also computes and transfers the KV of the unique text.
As a comparison, we evaluate the prefill time of this request when no global context cache is available. In this baseline, the KV of the full input (including the context and the unique text) is recomputed.
We evaluate the context lengths of 500, 2500, and 4500 respectively (so the total input lengths are 1k, 3k, and 5k).

Figure~\ref{fig:eval:prefill-efficiency} demonstrates \sys{}'s ability to leverage global context cache for KV migration between engines, significantly reducing prefill computation time. 
With context cache, KV migration ensures that only the unique text of 500 tokens requires computation, as opposed to prefilling the entire input.
For a total input length of 1000 tokens, KV migration effectively halves the prefill length, resulting in 1.7$\times$ speedup in prefill time.
Despite the increasing load of attention computation with longer input contexts, the prefill time with KV migration shows only a slight increase.
In contrast, prefill time without KV migration (i.e. with recomputation) grows linearly as input length increases from 1000 to 5000 tokens. This growing disparity underscores the importance of global context reuse and KV migration in \sys{}.

\begin{figure}[t]
    \centering
    \includegraphics[width=0.3\textwidth]{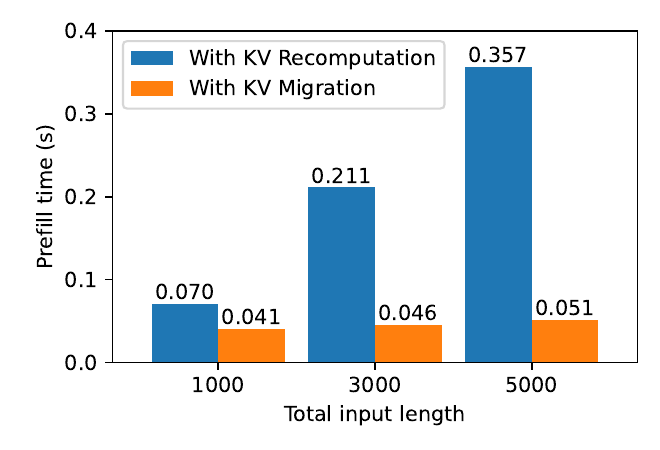}
    \caption{Llama3.1 8B prefill time comparison between ``with KV recomputation'' and ``with KV migration'' implemented in \sys{}. Context lengths are 500/2500/4500 respectively, and the length of unique text is 500 tokens. KV migration keeps the prefill time at a low level, compared to the prefill time that recomputes the full KV of the input.}
    \vspaceaftercap
    \label{fig:eval:prefill-efficiency}
\end{figure}

To further assess KV transfer efficiency during prefill, we evaluate transfer times under the same context cache settings. 
Table~\ref{table:eval:kv-transfer} presents the prefill and KV transfer times per model layer in Llama3.1 8B.
\sys{} overlaps the KV transfer with the prefill computation by employing the NVSHMEM library for GPU communication.
As the total input length increases from 1000 to 5000, the KV transfer time overlap ratio rises from 15.8\% to 55.4\%.
This increase occurs because the effective prefill length remains constant at 500 tokens (the length of unique text), while the size of transferred KV data grows linearly.
Notably, \sys{}'s use of one-sided NVSHMEM primitives allows KV transfer to overlap with ongoing decode computations on the receiver engine, minimizing communication overhead.

\begin{table}[!t]
    \centering
    \caption{Per-layer prefill time and KV transfer time of Llama3.1 8B. In \sys{}, KV transfer fully overlaps with prefill computation as long as the transfer time does not exceed the computation time.}
    \label{table:eval:kv-transfer}
    \begin{tabular}{l|c|c|c}
    \toprule
    \textbf{Input Length} & \textbf{1000}  & \textbf{3000} & \textbf{5000} \\
    \midrule
    $T_\mathrm{per-layer}$ (ms)   & 1.247  & 1.391  & 1.564  \\
    $T_\mathrm{KV-transfer}$ (ms) & 0.197  & 0.533  & 0.867  \\
    Transfer Time Ratio           & 15.8\% & 38.3\% & 55.4\% \\
    \bottomrule
    \end{tabular}
\end{table}

\subsection{Impact of PD Balance Ratio}
\label{sec:eval:balance-ratio}

\begin{figure}[t]
    \centering
    \includegraphics[width=0.45\textwidth]{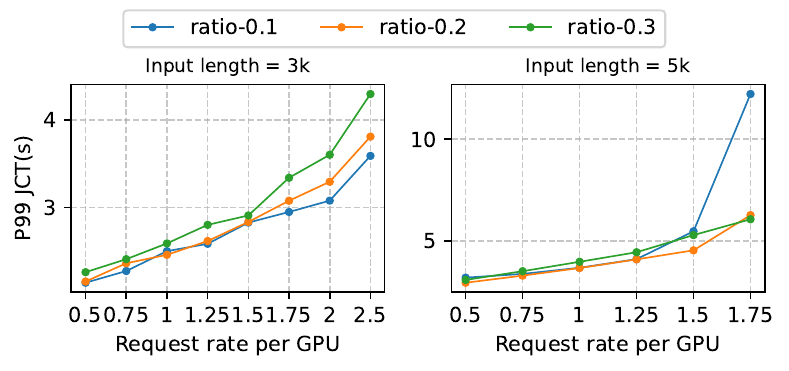}
    \caption{Impact of the PD balance ratio under different input lengths. Longer input requires higher PD balance ratio in order to further reduce the prefill engine pressure.}
    \vspaceaftercap
    \label{fig:eval:ablation}
\end{figure}

We have observed the performance benefit of the 1P1D-balance disaggregation pattern in \S\ref{sec:eval:disagg-patterns}.
In this section, we study the impact of different balance ratios that decide how much prefill workload will be transferred from the prefill engine to the decode engine.
We tested the balance ratios of 0.1, 0.2 and 0.3 under input lengths 3000 and 5000.
Results are shown in Figure~\ref{fig:eval:ablation}.
For the input length of 3000, the balance ratio of 0.1 offers up to 1.19 $\times$ speedup in P99 JCT than higher ratios.
When the input length is increased to 5000, it is more beneficial to transfer more prefill workload to decode engine,
especially in the scenario of a high request rate.
This is because a high request rate aggravates the pressure on the prefill engine.
As a result, requests can easily queue up on the prefill engine, and this congestion significantly slows down TTFT.
A higher balance ratio helps better alleviate the congestion and bring the system back to a prefill-decode balanced state.

\section{Related Work}
\label{sec:related-work}


\paragraph{Disaggregated Serving}
Recent advancements in LLM serving have focused on disaggregating the prefill and decode phases to optimize performance. SplitWise~\cite{patel2024splitwise} introduced a disaggregated inference approach that separates prefill and decode operations. DistServe~\cite{distserve} presented a distributed serving system that leverages disaggregation to improve goodput. TetriServe~\cite{hu2024inferenceinterferencedisaggregatellm} further refined this approach by introducing adaptive resource scheduling in disaggregated LLM serving. P/D-Serve~\cite{jin2024pdserveservingdisaggregatedlarge} dynamically adjusts P/D ratios and forward queued prefill for better performance.
LoongServe~\cite{wu2024loongserve} elastically adjusts the degree of parallelism to quickly serve variable-length requests in prefill and decode phases.
\sys{} complements these works by providing a simple yet effective programming model to support and dynamically reconfigure these disaggregated strategies. 

\paragraph{KV Cache Optimization}
\cite{wu2024layercondensedkvcacheefficient} proposed Layer-Condensed KV Cache, which computes and caches KVs for only a small number of layers, reducing memory consumption while maintaining competitive performance.
vLLM~\cite{vllm} developed PagedAttention, a transparent cache management layer that minimizes GPU memory fragmentation, improving inference speed.
Infinite-LLM~\cite{lin2024infinitellmefficientllmservice} extended PagedAttention to enable distributed deployment across servers.
Our system can leverage some of these optimizations and bring them to microserving scenarios.

\paragraph{Context Caching}
Context caching has emerged as a powerful technique to improve LLM serving performance. CacheGen~\cite{10.1145/3651890.3672274} introduced a novel approach to compress the KV cache and optimize its streaming, addressing bandwidth limitations in fetching large caches. EPIC~\cite{hu2024epicefficientpositionindependentcontext} presented a position-independent context caching system that enables modular KV cache reuse regardless of token chunk position. MemServe~\cite{hu2024memserve} introduced an elastic memory pool (MemPool) that manages distributed memory and KV caches across serving instances, combining context caching with disaggregated inference. \sys{} brings fine-grained API and flexible programming model that can enable effective combination and reconfiguration of the  context caching and disaggregated inference  strategies.

\paragraph{LLM Decoding Techniques}

Besides context caching techniques that optimize for prefill stage performance,
many LLM decoding techniques improve the decode stage efficiency by breaking the memory-bound auto-regressive decoding into local batch processing,
in order to benefit from the batching effects of LLMs.
Speculative decoding~\cite{chen-spec-decode,Leviathan-spec-decode,cai2024medusasimplellminference,li2024eaglespeculativesamplingrequires} employs a small draft model to provide multiple draft tokens at a time for batched verification.
Sarathi-Serve~\cite{agrawal2024taming} introduces chunked-prefills and stall-free scheduling to optimize the throughput-latency tradeoff in LLM inference.
With engine-level support, \sys{} can benefit from these techniques in performance and bring them to microserving without any router-level changes.

\section{Conclusion}
\label{sec:conclusion}

We introduce \sys{}, a multi-level architecture for structuring and programming LLM inference services.
 By providing fine-grained REST APIs, and a unified KV cache interface, \sys{} allows easy implementation and customization of various disaggregation strategies while maintaining competitive performance.
We hope this work will encourage additional studies of dynamic reconfiguration and orchestration 
strategies in microserving LLM workloads.

\bibliography{paper}
\bibliographystyle{mlsys2024}


\end{document}